\documentclass[journal]{IEEEtran}
\usepackage[dvips]{graphicx}
\graphicspath{{../eps/}}
\DeclareGraphicsExtensions{.eps}

\begin{document}

\title{Distributed Arithmetic Coding for Sources with Hidden Markov Correlation}

\author{
Yong~Fang~and~Jechang~Jeong,~\IEEEmembership{Member,~IEEE}
\thanks{
This research was supported by Seoul Future Contents Convergence (SFCC) Cluster established by Seoul R\&BD Program.}
\thanks{The authors are with the Lab. of ICSP, Dep. of Electronics and Communications Engineering, Hanyang University, Haengdang-dong, Seongdong-gu, Seoul 133-791, Korea
(e-mail: yfang79@gmail.com).}
}

\maketitle

\begin{abstract}
Distributed arithmetic coding (DAC) has been shown to be effective for Slepian-Wolf coding, especially for short data blocks. In this letter, we propose to use the DAC to compress momery-correlated sources. More specifically, the correlation between sources is modeled as a hidden Markov process. Experimental results show that the performance is close to the theoretical Slepian-Wolf limit.
\end{abstract}

\begin{IEEEkeywords}
Distributed Source Coding (DSC), Slepian-Wolf Coding (SWC), Distributed Arithmetic Coding (DAC), Hidden Markov Correlation, Forward Algorithm.
\end{IEEEkeywords}

\section{Introduction}
\IEEEPARstart{W}{e} consider the problem of Slepian-Wolf Coding (SWC) with decoder Side Information (SI). The encoder compresses discrete source $X=\{x_t\}_{t=1}^{N}$ in the absence of $Y=\{y_t\}_{t=1}^{N}$, discretely correlated SI. Slepian-Wolf theorem points out that lossless compression is achievable at rates $R \geq H(X|Y)$, the conditional entropy of $X$ given $Y$, where both $X$ and $Y$ are discrete random processes [1]. Conventionally, channel codes, such as turbo codes [2] or Low-Density Parity-Check (LDPC) codes [3], are used to deal with the SWC problem.

Recently, some SWC techniques based on entropy coding are proposed, such as Distributed Arithmetic Coding (DAC) [4, 5] and Overlapped Quasi-Arithmetic Coding (OQAC) [6]. These schemes can be seen as an extension of classic Arithmetic Coding (AC) whose principle is to encode source $X$ at rates $H(X|Y) \leq R < H(X)$ by allowing overlapped intervals. The overlapping leads to a larger final interval and hence a  shorter codeword. However, ambiguous codewords are unavoidable at the same time. A soft joint decoder exploits SI $Y$ to decode $X$. Afterwards, the time-shared DAC (TS-DAC) [7] is proposed to deal with the symmetric SWC problem. To realize rate-incremental SWC, the rate-compatible DAC is proposed in [8].

In this paper, we research how to use the DAC to compress sources with hidden Markov correlation.

\section{Binary Distributed Arithmetic Coding}
Let $p$ be the bias probability of binary source $X$, i.e., $p = P(x_t = 1)$. In the classic AC, source symbol $x_t$ is iteratively mapped onto sub-intervals of $[0, 1)$, whose lengths are proportional to $(1-p)$ and $p$. The resulting rate is $R \geq H(X)$. In the DAC [4], interval lengths are proportional to the modified probabilities $(1-p)^\gamma$ and $p^\gamma$, where
\begin{equation}
\frac{H(X|Y)}{H(X)} \leq \gamma \leq 1.
\end{equation}
The resulting rate is $R \geq \gamma H(X) \geq H(X|Y)$. To fit the $[0, 1)$ interval, the sub-intervals have to be partially overlapped. More specifically, symbols $x_t = 0$ and $x_t = 1$ correspond to intervals $[0, (1-p)^\gamma)$ and $[1-p^\gamma, 1)$, respectively. It is just the overlapping that leads to a larger final interval, and hence a shorter codeword. However, as a cost, the decoder can not decode $X$ unambiguously without $Y$.

To describe the decoding process, we define a ternary symbol set $\{0, \chi, 1\}$, where $\chi$ represents a decoding ambiguity. Let $C_X$ be the codeword and $\tilde{x}_t$ be the $t$-th decoded symbol, then
\begin{equation}
\setlength{\nulldelimiterspace}{0pt}
\tilde{x}_t = \left\{
\begin{IEEEeqnarraybox} [\relax] [c] {l's}
0, &$0 \leq C_X < 1 - p^\gamma$ \\
\chi, &$1 - p^\gamma \leq C_X < (1-p)^\gamma$\\
1, &$(1-p)^\gamma \leq C_X < 1$%
\end{IEEEeqnarraybox}.
\right.
\end{equation}

When the $t$-th symbol is decoded, if $\tilde{x}_t = \chi$, the decoder performs a branching: two candidate branches are generated, corresponding to two alternative symbols $x_t = 0$ and $x_t = 1$. For each new branch, its metric is updated and the corresponding interval is selected for next iteration. To reduce complexity, every time after decoding a symbol, the decoder uses the $M$-algorithm to keep at most $M$ branches with the best partial metric, and prunes others [4].

Note that the metric is not reliabe for the very last symbols of a finite length sequence $X$ [5]. This problem is solved by encoding the last $T$ symbols without interval overlapping [5]. It means that for $1 \leq t \leq (N-T)$, $x_t$ is mapped onto $[0, (1-p)^\gamma)$ and $[1-p^\gamma, 1)$; while for $(N-T+1) \leq t \leq N$, $x_t$ is mapped onto $[0, 1-p)$ and $[1-p, 1)$.

Therefore, a binary DAC system can be described by four parameters: $\{p, \gamma, M, T\}$.

\section{Hidden Markov Model and Forward Algorithm}
Let $S = \{s_t\}_{t=1}^N$ be a sequence of states and $Z = \{z_t\}_{t=1}^N$ be a sequence of observations. A hidden Markov process is defined by $\lambda = (A, B, \pi)$:

$A=\{a_{ji}\}$: state transition probability matrix, where $a_{ji} = P(s_t = i | s_{t-1} = j)$;

$B=\{b_i(k)\}$: observation probability distribution, where $b_i(k) = P(z_t = k | s_t = i)$;

$\pi=\{\pi_{i}\}$: initial state distribution, where $\pi_i = P(s_1 = i)$.

The aim of forward algorithm is to compute $P(z_1, ..., z_t|\lambda)$, given observation $\{z_1, ..., z_t\}$ and model $\lambda$.
Let $\alpha_t(i)$ be the probability of observing the partial sequence $\{z_1, ..., z_t\}$ such that state $s_t$ is $i$, i.e.,
\begin{equation}
\alpha_t(i) = P(z_1, ..., z_t, s_t = i | \lambda).
\end{equation}
Initially, we have
\begin{equation}
\alpha_1(i) = \pi_i b_i(z_1).
\end{equation}
For $t>1$, $\alpha_t(i)$ can be induced through iteration
\begin{equation}
\alpha_{t}(i) = \{ \sum_j [\alpha_{t-1}(j) a_{ji}] \} b_{i}(z_t).
\end{equation}
Therefore,
\begin{equation}
P(z_1, ..., z_t|\lambda) = \sum_i {\alpha_t(i)}.
\end{equation}

In practice, $\alpha_t(i)$ is usually normalized by
\begin{equation}
\alpha_t(i) = \frac {\alpha_t(i)} {\delta_t},
\end{equation}
where
\begin{equation}
\delta_t = \sum_i {\alpha_t(i)}.
\end{equation}
In this case, we have
\begin{equation}
P(z_1, ..., z_t|\lambda) = \prod _{t'=1} ^{t} {\delta_{t'}}.
\end{equation}

\section{DAC for Hidden Markov Correlation}
Assume that binary source $X$ and SI $Y$ are correlated by $Y = X \oplus Z$, where $Z$ is generated by a hidden Markov model with parameter $\lambda$. $X$ is encoded using a $\{p, \gamma, M, T\}$ DAC encoder. The decoding process is very similar to what described in [4]. The only difference is that the forward algorithm is embedded into the DAC decoder and the metric of each branch is replaced by $P(z_1, ..., z_t|\lambda)$, where $z_t = x_t \oplus y_t$.

\section{Experimental Results}
We have implemented a 16-bit DAC codec system. The bias probability of $X$ is $p=0.5$. According to the recommendation of [5], we set $M=2048$ and $T=15$. The same 2-state (0 and 1) and 2-output (0 and 1) sources as in [9] are used in simulations (see Table I). The length of data block used in each test is $N=1024$.

To achieve lossless compression, each test starts from $\gamma=H(X|Y)$ (see Table II). If the decoding fails, we increase $\gamma$ with 0.01. Such process is iterated until the decoding succeeds. For each model, results are averaged over 100 trials. Experimental results are enlisted in Table II.

For comparison, also included in Table II are experimental results for the same settings from [9]. In each test of [9], $N=16384$ source symbols are encoded using an LDPC code. In addition, to synchronize the hidden Markov model, $N\alpha$ original source symbols are sent to the decoder directly without compression.

The results show that the DAC performs similarly to or slightly better than (for models 1 and 2) the LDPC-based approach [9]. Moreover, for hidden Markov correlation, the DAC outperforms the LDPC-based approach in two aspects:

1). The LDPC-based approach requires longer codes to achieve better performance, while the DAC is insensitive to code length [4].

2). For the LDPC-based approach, to synchronize the hidden Markov model, a certain proportion of original source symbols must be sent to the decoder as ``seeds". However, it is hard to determine $\alpha$, the optimal proportion of ``seeds". The results reported in [9] were obtained through an exhaustive search, which limits its application in practice.

\begin{table}[!t]
\renewcommand{\arraystretch}{1.3}
 
\caption{Models for Simulation}
\centering
\begin{tabular}{c||c}
\hline
\bfseries $model$ & \bfseries $\{a_{00}, a_{11}, b_0(0), b_1(1)\}$\\
\hline\hline
1 & \{0.01, 0.03, 0.99, 0.98\}\\
\hline
2 & \{0.01, 0.065, 0.95, 0.925\}\\
\hline
3 & \{0.97, 0.967, 0.93, 0.973\}\\
\hline
4 & \{0.99, 0.989, 0.945, 0.9895\}\\
\hline
\end{tabular}
\end{table}

\begin{table}[!t]
\renewcommand{\arraystretch}{1.3}
\caption{Experimental Results}
\centering
\begin{tabular}{c||c||c||c}
\hline
\bfseries $model$ & \bfseries $H(X|Y)$ & \bfseries [9] & \bfseries DAC\\
\hline\hline
1 & 0.24 & 0.36 & 0.345908\\
\hline
2 & 0.52 & 0.67 & 0.648594\\
\hline
3 & 0.45 & 0.58 & 0.585645\\
\hline
4 & 0.28 & 0.42 & 0.427236\\
\hline
\end{tabular}
\end{table}

\section{Conclusion}
This paper researches the compression of sources with hidden Markov correlation using the DAC. The forward algorithm is incorporated into the DAC decoder. The results are similar to that of the LPDC-based approach. Compared to the LDPC-based approach, the DAC is more suitable for practical applications.

\end{document}